\documentclass[superscriptaddress,prb,twocolumn]{revtex4} 
\usepackage{graphicx,anysize}
\usepackage{amssymb}

\begin{document}
\title{Giant direct magnetoelectric effect in strained multiferroic heterostructures}
\author{Pierre-Eymeric Janolin}
  \email{Pierre-Eymeric.Janolin@ecp.fr}
\affiliation{Laboratoire Structures, Propri\'et\'es et Mod\'elisation des Solides, UMR CNRS-\'Ecole Centrale Paris,
Grande Voie des Vignes, 92295 Ch\^atenay-Malabry Cedex, France}
\author{Nikolay A. Pertsev}
\affiliation{A. F. Ioffe Physico-Technical Institute, Russian Academy of Sciences, 194021 St. Petersburg, Russia}
\affiliation{Laboratoire Structures, Propri\'et\'es et Mod\'elisation des Solides, UMR CNRS-\'Ecole Centrale Paris,
Grande Voie des Vignes, 92295 Ch\^atenay-Malabry Cedex, France}
\author{David Sichuga}
\affiliation{Physics Department, Augusta Technical College, Augusta, Georgia, 30906, USA}
\author{L. Bellaiche}
\affiliation{Institute for Nanoscience and Engineering and Physics Department, University of Arkansas, Fayetteville, Arkansas 72701, USA}

\begin{abstract}
The direct magnetoelectric (ME) effect mediated by lattice strains induced in a ferroelectric film by a ferromagnetic substrate is evaluated using first-principles-based calculations. To that end, the strain sensitivity of ferroelectric polarization and the film permittivity are calculated as a function of the in-plane biaxial strain for Pb(Zr$_{0.52}$Ti$_{0.48}$)O$_3$ films under various depolarizing fields. It is found that the ME voltage coefficient varies nonmonotonically with this strain and may reach giant values exceeding 100~V\,cm$^{-1}$\,Oe$^{-1}$ over a strain range that can be controlled through the electrical boundary conditions. 
\end{abstract}	

\date{\today}	
\maketitle
Multiferroic materials exhibiting magnetoelectric (ME) effects represent the subject of cut\-ting-edge research since they are promising for applications in the next generation of microelectronic and nanoelectronic devices \cite{Vaz}. As the room-temperature ME effects in single-phase materials are weak, the theoretical and experimental studies are currently focused on composite multiferroics. Here the interface-related and proximity effects may strongly increase the ME response, as found, in particular, for the strain-mediated ME effects in ferroelectric-ferromagnetic hybrids \cite{Nan}.

The design of advanced multifunctional materials may be greatly facilitated by theoretical modeling. 
In particular, 
\textit{ab initio} and first-principles-based calculations already proved to be able to predict physical properties of single crystals and thin films directly from microscopic phenomena occurring at the atomic scale \cite{RefsFP}. Application of first-principles calculations to ferroelectric-ferromagnetic heterostructures and composites, however, is a challenging task as the behavior of such systems is much more complex than that of single-phase materials. Recently, several studies were performed on multiferroic heterostructures with the 2-2 connectivity \cite{Tsymbal2006, Tsymbal2008, Tsymbal2009, Demkov2009, Mertig2010}. However, the predicted ME effects are either weak or extremely localized at the interfaces, which limit their possible applications in microelectronic devices. In contrast, the \textit{strain-mediated} ME effect spans over the whole volume of the active phase in a properly designed heterostructure. Accordingly, the macroscopic ME response becomes strong as well, which is critical for the development of ME sensors of magnetic fields and many other devices \cite{Nan}.  

Theoretical predictions of the strain-mediated direct ME effect generally require the determination of three ingredients: (i)  the deformations induced in the ferromagnetic component by the applied magnetic field; (ii) the degree of strain transmission through the interface between two ferroic constituents of a hybrid material system; and (iii) the response of polarization and dielectric susceptibility of the ferroelectric phase to changes in lattice strains. In the case of epitaxial heterostructures, the mechanical coupling between two phases is usually very strong so that the strain transmission may be perfect. The deformation response of a ferromagnet to the measuring ac magnetic field $\delta$\textbf{H} becomes large only in the presence of a bias magnetic field \textbf{H} \cite{Moffett, Wang, Clark, Dong}. The dependence of this response on \textbf{H} is nonmonotonic and cannot be calculated theoretically with sufficient accuracy \cite{NT2010}. Accordingly, the first ingredient should be evaluated using the available experimental data on the field dependence of magnetostrictive deformations. On the contrary, the properties of a single-crystalline ferroelectric phase subjected to strains can be calculated from first principles with a high accuracy \cite{Review FP}.

In this Letter, we calculate the strain-mediated direct ME effect for a multiferroic hybrid in the form of a thin ferroelectric film sandwiched between two electrodes and coupled to a thick ferromagnetic substrate. This geometry maximizes the ME coefficients since the film does not suppress the magnetic-field-induced deformations of the substrate. As a representative example of a ferroelectric material, we have chosen the Pb(Zr$_{0.52}$Ti$_{0.48}$)O$_3$ (PZT) disordered solid solution with the morphotropic composition. The strain sensitivity of the out-of-plane polarization and the film permittivity are calculated for (001)-oriented PZT films 
by first-principles-based methods. Since epitaxial films may be strained to a different extent during the deposition, these characteristics are calculated as a function of initial biaxial in-plane strain $\eta_m$ typically induced in perovskite films deposited on (001)-oriented cubic substrates. The bottom electrode is assumed to be fully strained by the substrate or thick buffer layer due to coherent lattice matching at the interface. The presence of electrodes is taken into account in the calculations through the $\beta$ parameter that describes partial screening of the depolarizing field \cite{beta}. The film is modeled by a 12x12x12 supercell with periodic boundary conditions along the [100] and [010] pseudocubic directions (chosen as the $x_1$ and $x_2$ axes, respectively). The nonperiodic [001] direction ($x_3$ axis) thus allows for a finite thickness of the film.

The effective Hamiltonian used in our calculations is described in Refs.\cite{Sichuga}. Importantly, it takes into account not only the ferroelectric local modes ($u_{i}$) ($i$ = 1,2,3) and strains ($\eta_{ij}$), but also rotations of the oxygen octahedra ($\omega_i$) as well as the fluctuations of $u_{i}$, $\eta_{ij}$, and $\omega_i$. This feature, together with the account of the depolarizing-field and finite-thickness-related effects, represent the major differences from the previous phenomenological calculations \cite{NT2010}. As a result, our study provides a significantly improved description of the ME effect, notably bringing into light the crucial role of the depolarizing field. 

Mechanical boundary conditions imposed on a thin epitaxial film involve three fixed components of the strain tensor, which for the cube-on-cube epitaxy become $\eta_{11}=\eta_{22}=\eta_m$ and $\eta_{12}=0$. The biaxial misfit strain $\eta_m$ equals $\eta_m=a_{film}/a_{ref}-1$, where $a_{film}$ is the actual in-plane lattice parameter of the film and $a_{ref}$=4.065\AA\ is the lattice constant of the prototypic cubic phase of bulk PZT at room temperature \cite{PZTbulk}. It is worth noting that $a_{film}$ coincides with the lattice parameter of the substrate in the case of coherent epitaxy, but may differ from it considerably in hybrids with misfit dislocations. 

The total  energy of the supercell was used in Monte-Carlo (MC) simulations with 4.10$^4$ MC sweeps. To compute the strain-mediated ME effect, the code was modified to run calculation as a function of strain at constant temperature (300\,K). 

\begin{figure}[ht]
  \centering
   \includegraphics[angle=-90,width=.45\textwidth]{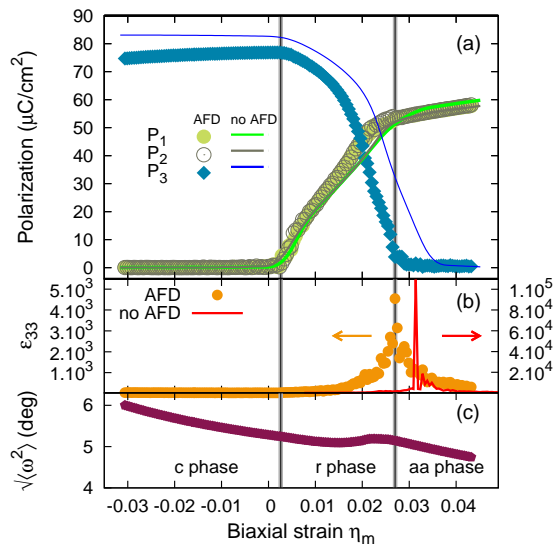}
 \caption{Polarization components (a), out-of-plane dielectric constant (b) and average magnitude of the fluctuations of the \emph{local} oxygen octahedra tilting in each unit cell (c) as a function of biaxial misfit strain for a 48\AA-thick PZT film at 300 K under partially screened depolarizing field ($\beta=0.98$). The influence of oxygen octahedra rotations (labelled AFD) is illustrated on panels (a) and (b).}
 \label{fig:biaxial}
\end{figure}

Figure \ref{fig:biaxial} displays the predicted effect of biaxial strain on the polarization components $P_i$ calculated using the standard relationship between $P_i$ and $u_i$
and the out-of-plane dielectric constant $\epsilon_{33}$ under partially screened depolarizing field ($\beta$=0.98).
With the strain $\eta_m$ varying from compressive to tensile, the equilibrium ferroelectric phase changes from the tetragonal $c$ one (space group \textit{P}4\textit{mm}) to the monoclinic  $r$ phase (\textit{Cm})  and then to the orthorhombic $aa$ phase (\textit{Cmm}2). The permittivity $\epsilon_{33}$ strongly increases at the $r$-$aa$ phase transition because 
$P_3$ goes to zero here.
The account of rotations of oxygen octahedra changes the film polarization behavior dramatically, despite the fact that the mean angle ($\langle\omega\rangle$) is found to be zero, in agreement with ref.\cite{Sichuga}. Indeed, the decrease of $P_3$ with increasing compressive strain is linked to the average magnitude of the fluctuations of the local oxygen octahedra tilting in each unit cell (see the evolution of $\sqrt{\langle\omega^2\rangle}$ with $\eta_m$ on Fig.\ref{fig:biaxial}(c)), as when rotations are frozen $P_3$ increases slighty (see Fig.\ref{fig:biaxial}(a)). In addition, these rotations change the 
position of the $r$-$aa$ phase transition and associated dielectric peak and affect the maximum value of permittivity

The film strain states leading to the formation of various phases shown in Fig.\ref{fig:biaxial} can be achieved through an appropriate buffer layer deposited on a ferromagnetic substrate. When the magnetic field \textbf{H} is turned on, the film strains change due to substrate magnetostrictive deformations. In the symmetric case, where \textbf{H} is orthogonal to the film plane and the substrate has a four-fold symmetry about the $x_3$ axis, these changes reduce to a variation of the biaxial strain $\eta_m$.  Hence the ME effect can be evaluated  via the strain sensitivity $\partial P_3/\partial \eta_m$ of the out-of-plane polarization $P_3$ with respect to $\eta_m$. 

\begin{figure}[ht]
\centering
  \includegraphics[angle=-90,width=.45\textwidth]{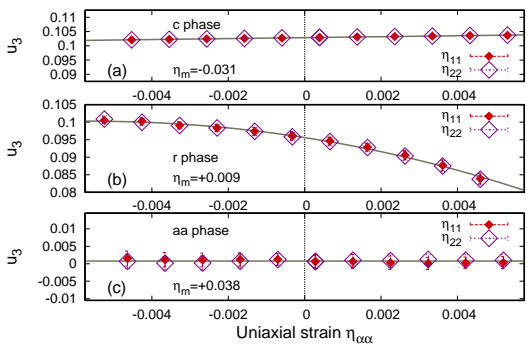}
 \caption{Out-of-plane local mode $u_3$ as a function of uniaxial strain $\eta_{11}$ or $\eta_{22}$ (closed and open symbols, respectively) in the (a) $c$, (b) $r$, and (c) $aa$ phases. The lines represent the best fits with a linear ((a) and (c)) and power-law (b) functions.}
 \label{fig:Unimodes}
\end{figure}

The magnetic field parallel to the film surfaces, however, breaks the initial isotropy of in-plane strains. To quantify the ME effect appearing in this situation, one has to calculate the sensitivities of $P_3$ with respect to uniaxial strains $\eta_{11}$ and  $\eta_{22}$. Therefore, we computed the evolution of the local mode $u_3$ as a function of uniaxial strain $\eta_{\alpha\alpha}$ superimposed on a biaxial strain $\eta_m$. Representative results obtained for the three ferroelectric phases are shown in Fig.\ref{fig:Unimodes}. It is seen that $\eta_{11}$ and $\eta_{22}$ have the same effect on $u_3$, which is consistent with the symmetry of the phases. In the $c$ and $aa$ phases, the calculated dependences are linear. In contrast, the evolution of the local mode occurring in the $r$ phase follows a power law. However, on the scale of small strains induced by the measuring magnetic field $\delta$\textbf{H} the linear approximation proposed in Ref.\cite{NT2010} holds. It should be noted that the dependences shown in Fig.\ref{fig:Unimodes} also enable us to allow for the influence of the bias magnetic field \textbf{H} (applied along the [100] or [010] direction). The strain changes induced by \textbf{H}, however, usually may be neglected in comparison with initial strains because relevant magnetostrictive deformations are well below 10$^{-3}$ \cite{Moffett, Wang, Clark, Dong}.

\begin{figure}[ht]
  \centering
  \includegraphics[angle=-90,width=.45\textwidth]{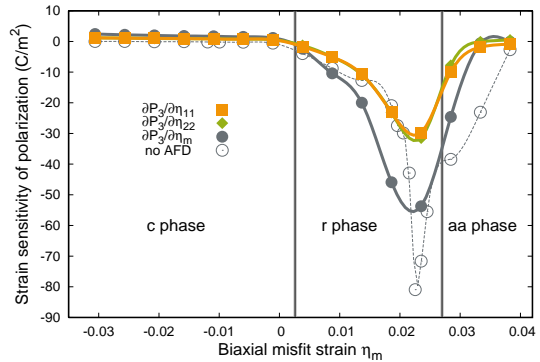}
 \caption{Sensitivities of the out-of-plane polarization to $\eta_{11}$ (squares), $\eta_{22}$ (diamonds), and $\eta_m$ (with AFD full circles, without AFD empty ones) plotted as a function of the initial biaxial strain $\eta_m$. The lines are cubic spline fits to the data points.}
 \label{fig:strsen}
\end{figure}

The strain sensitivities $\partial P_3/\partial \eta_{\alpha\alpha}$ calculated from the local mode $u_3$ are plotted in Fig.\ref{fig:strsen} as a function of the initial biaxial strain. As expected from the symmetries of predicted phases, within the accuracy of calculations $\partial P_3/\partial \eta_{11}$ = $\partial P_3/\partial \eta_{22}$. At the same time, these sensitivities are significantly smaller than the sensitivity $\partial P_3/\partial \eta_m$ of polarization with respect to the biaxial strain itself (with and without AFD), which is also plotted in Fig.\ref{fig:strsen}. This result is consistent with the prediction of the thermodynamic theory \cite{NT2010} giving $\partial P_3/\partial \eta_m$ = 2 $\partial P_3/\partial \eta_{11}$ owing to the superposition principle. The positions of the maxima, however, are shifted away from the $r$-$aa$ transition, in contrast to thermodynamic theory. This is caused by inhomogeneous strains arising at the local scale within the film because strains are calculated in each cell of the supercell [15]. 

\begin{figure}[ht]
  \centering
  \includegraphics[angle=-90,width=.45\textwidth]{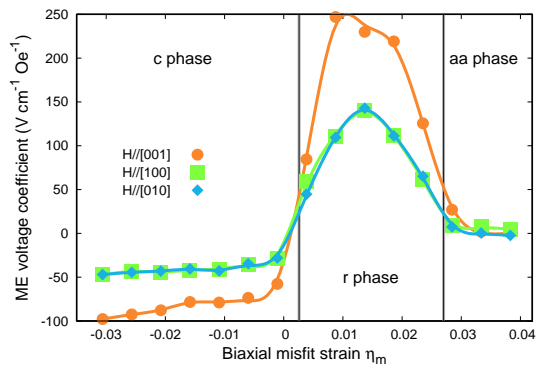}
 \caption{Magnetoelectric voltage coefficient of a multiferroic hybrid comprising a 48\AA-thick PZT film coupled to a ferromagnetic substrate subjected to a magnetic field parallel to the [001] (bullets), [100] (squares), or [010] (diamonds) crystallographic axes of the film.The lines are cubic spline fits to the data points.}
 \label{fig:MEVC}
\end{figure}

The calculated strain sensitivities and dielectric constant enable us to evaluate the polarization ($\alpha_{P3j}$=$\partial P_3/\partial H_j$) and voltage ($\alpha_{E3j}$=$\partial E_3/\partial H_j$) ME coefficients characterizing the discussed multiferroic hybrid. Since the figure of merit of a magnetic field sensor is the output voltage it delivers, we focus on the ME voltage coefficients governing the electric field $E_3$ induced in the ferroelectric parallel-plate capacitor. If the magnetic field is parallel to the [100] or [010] crys\-tal\-lo\-gra\-phic axis of the film and does not induce shear strains here, we have $\alpha_{E3j}$=$-\frac{1}{\epsilon_{33}}\left(\frac{\partial P_3}{\partial \eta_{11}}d^{m}_{j11}+\frac{\partial P_3}{\partial \eta_{22}}d^{m}_{j22}\right)$,
%
%
where $d^{m}_{j\alpha\alpha}$ are the substrate piezomagnetic coefficients at a given bias magnetic field \cite{NT2010}. Taking the maximum longitudinal coefficient $d^{m}_{111}$=$d^{m}_{222}$=5\,10$^{-8}$ m\,A$^{-1}$ as measured for the FeBSiC alloy (see Fig. 3 in ref. \cite{Dong}) and estimating the transverse coefficient to be $d^{m}_{122}$=\-$d^{m}_{211}$=-2.5\,10$^{-8}$\,m A$^{-1}$ from the condition of volume conservation \cite{PRB2009}, we calculated the ME voltage coefficients $\alpha_{E3j}$ as a function of initial strain $\eta_m$. Figure \ref{fig:MEVC} shows that within the accuracy of calculations $\alpha_{E31}$=$\alpha_{E32}$ and further demonstrates a nonmonotonic strain dependence of these transverse ME coefficients. 
Remarkably, they reach a giant value of $\sim$150~V\,cm$^{-1}$\,Oe$^{-1}$ in the middle of the stability range of the $r$ phase, which significantly corrects the prediction of the thermodynamic theory \cite{PRB2009, NT2010}. Moreover, $\alpha_{E31}$ and $\alpha_{E32}$ change sign near the $c$-$r$ phase transition, i.e. where $\partial P_3/\partial \eta_{\alpha\alpha}$ does. Accordingly, the ME response vanishes at a certain misfit strain so that even the presence of out-of-plane polarization does not guarantee the existence of a non-zero ME effect. 

The longitudinal ME coefficient $\alpha_{E33}$ can be evaluated from the relation $\alpha_{E33}$=$-\frac{1}{\epsilon_{33}}\frac{\partial P_3}{\partial \eta_m}d^{*m}_{311}$, where $d^{*m}_{311}$ is the effective piezomagnetic coefficient characterizing local magnetostrictive deformations beneath the film/substrate interface. Since the substrate magnetization here is orthogonal to the interface, the demagnetizing field is not negligible, in contrast to a plate-like substrate subjected to an in-plane magnetic field, which was considered above. As a result, the relevant local field may be much smaller than the applied field, which makes $d^{*m}_{311}<<d^{m}_{122}$ for a plate-like substrate. However, the effective coefficient $d^{*m}_{311}$ can be increased strongly by employing a rod-like substrate \cite{NT2010}. 
Since $\alpha_{E33}$ is not very sensitive to $d^{*m}_{311}$, we plotted this coefficient in Fig.\ref{fig:MEVC} assuming that $d^{*m}_{311}$ = $d^{m}_{122}$. Although this plot represents the upper bound of $\alpha_{E33}$, we see that the longitudinal ME coefficient may reach giant values about 200~V\,cm$^{-1}$\,Oe$^{-1}$ even if $d^{*m}_{311}$ is 20\% smaller than $d^{m}_{122}$.

Finally, we studied how the ME performance of the device depends on the screening ability of electrodes. This effect can be probed by changing the parameter $\beta$ that governs the amount of screened depolarizing field \cite{beta}. Although formally $\beta$ may range from unity to zero, for a device with metallic electrodes only values close to unity are relevant. Moreover, since the calculations showed that at $\beta<0.95$ the monoclinic $r$ phase cannot be stabilized at room temperature, we restricted our analysis to $0.95\leq\beta\leq 1.00$. It should be noted that $\beta$ can be estimated from the relation $\beta=1-\epsilon_0/(\epsilon_0+c_it)$, where $c_i$ is the capacitance density associated with two film-electrode interfaces, $\epsilon_0$ is the permittivity fo the vacuum, and $t$ is the film thickness. Using the values of $c_i$ determined by first-principles calculations \cite{Stengel2009}, we find that typical ferroelectric capacitors with Pt and SrRuO$_3$ electrodes have $\beta$ exceeding 0.99 at the considered thickness $t$=4.8\,nm.

\begin{figure}[ht]
  \centering
  \includegraphics[angle=-90,width=0.45\textwidth]{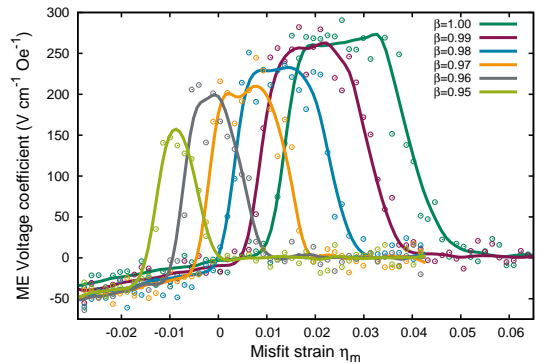}
 \caption{Longitudinal ME voltage coefficient calculated under different electrical boundary conditions varying from the perfect screening of the depolarizing field ($\beta$=1.00) to an imperfect screening with $\beta$=0.95. The lines are cubic spline fits to the data points.}
 \label{fig:BetaMEVC}
\end{figure}

Figure \ref{fig:BetaMEVC} shows the longitudinal ME voltage coefficient calculated under different screening conditions. It can be seen that the highest ME response is achieved at zero depolarizing field ($\beta$=1.00). At the same time, the maximum value of the ME coefficient remains very high even at $\beta=0.95$. Remarkably, with decreasing screening ability of electrodes, the peak of ME response shifts from positive to negative misfit strains. This is in agreement with the effect of changing $\beta$ on the temperature-misfit strain phase diagram of BaTiO$_3$ \cite{Bo}. This feature provides a new route for the development of ultra-sensitive ME sensors since it shows that the ME response can be enhanced not only by tuning the misfit strains \cite{NT2010} but also by selecting an appropriate electrode material. 

In summary, we performed the first calculation of the strain-mediated ME effect by a first-principles-based technique. It was found that a multiferroic hybrid comprising a Pb(Zr$_{0.52}$Ti$_{0.48}$)O$_3$ thin film coupled to a ferromagnetic substrate may display giant longitudinal and transverse ME voltage coefficients exceeding 100~V\,cm$^{-1}$\,Oe$^{-1}$. Remarkably, this giant ME response exists in a wide strain range $\delta\eta_m$$\sim$0.01, and the ME peak can be positioned exactly at the initial biaxial strain $\eta_m$ existing in the film by an appropriate choice of electrodes.

The authors thank Igor Kornev for valuable discussions in the course of this work. D.S. and L.B. mostly acknowledge the financial support of NSF grant DMR-0701558 and NSF No. Grant DMR-1066158. ONR Grants N00014-11-1-0384 and N00014-08-1-0915, ARO Grant W911NF-12-1-0085 and the Department of Energy, Office of Basic Energy Sciences, under contract ER-46612, are also acknowledged for discussions with scientists sponsored by these grants. Some computations were also made possible thanks to the MRI grant 0959124 from NSF.

\end{document}